\documentclass[aps,prl,twocolumn,superscriptaddress]{revtex4}%
\usepackage{graphics}
\usepackage{amsmath}
\usepackage{graphicx}
\usepackage{amsfonts}
\usepackage{amssymb}
\usepackage{float}
\usepackage{longtable}
\usepackage{epsfig}
\usepackage{color}
\usepackage{latexsym}
\usepackage{theorem}
\usepackage{bbm}
\usepackage{bm}
\usepackage{verbatim}
\usepackage{psfrag}%
\DeclareMathOperator*{\Tr}{Tr}

\begin{document}
\title{Dense coding capacity of a quantum channel}
\author{Riccardo Laurenza}
\affiliation{QSTAR, INO-CNR and LENS, Largo Enrico Fermi 2, 50125 Firenze, Italy}
\author{Cosmo Lupo}
\affiliation{Department of Physics and Astronomy, University of Sheffield, Hounsfield Road,
Sheffield, S3 7RH, UK}
\author{Seth Lloyd}
\affiliation{Department of Mechanical Engineering, Massachusetts Institute of Technology,
Cambridge MA 02139, USA}
\affiliation{Research Laboratory of Electronics, Massachusetts Institute of Technology,
Cambridge MA 02139, USA}
\author{Stefano Pirandola}
\affiliation{Research Laboratory of Electronics, Massachusetts Institute of Technology,
Cambridge MA 02139, USA}
\affiliation{Department of Computer Science, University of York, York YO10 5GH, UK}

\begin{abstract}
We consider the fundamental protocol of dense coding of classical information
assuming that noise affects both the forward and backward communication lines
between Alice and Bob. Assuming that this noise is described by the same
quantum channel, we define its dense coding capacity by optimizing over all
adaptive strategies that Alice can implement, while Bob encodes the
information by means of Pauli operators. Exploiting techniques of channel
simulation and protocol stretching, we are able to establish the dense coding
capacity of Pauli channels in arbitrary finite dimension, with simple formulas
for depolarizing and dephasing qubit channels.

\end{abstract}
\maketitle

\textit{Introduction}.--~One of the most essential resources for quantum
comunication and information processing~\cite{NiCh,review,first,HolevoBOOK} is
represented by entanglement. Quantum entanglement describes correlations
outside the classical realm and it is at the core of the realization of many
quantum tasks, including quantum teleportation~\cite{tele1,tele2}, quantum
cryptography~\cite{BB84} and dense coding~\cite{BenDense}. The dense coding
protocol allows two parties to transmit classical information encoded on
quantum systems with the aid of shared entanglement. By employing a bipartite
entangled state, it is possible to encode $2\log_{2}d$ bits of classical
information in a $d$-dimensional system, thus overcoming the upper bound
$\log_{2}d$ on the unassisted classical capacity.

In ideal conditions, a dense coding scheme exploits a noiseless quantum
channel between Alice and Bob. Through this quantum channel, Alice sends to
Bob part $B$ of a bipartite entangled state $\sigma_{AB}$. Once received by
Bob, system $B$\ is subject to a Pauli operator $U_{x}$ with probability
$P_{x}$. The encoded system is sent back to Alice through the second use of
the noiseless quantum channel. At the output, Alice implements a joint quantum
measurement on $A$ and $B$ to retrieve the classical information. In this
case, the capacity $C(\sigma_{AB})$ is~\cite{Hiro,ZiBu}
\begin{equation}
C(\sigma_{AB})=\max\{\log_{2}d,~\log_{2}d+S(\sigma_{B})-S(\sigma_{AB})\},
\end{equation}
where $\sigma_{B}=\Tr_{A}\sigma_{AB}$ and $S(\sigma):=-\Tr(\sigma\log
_{2}\sigma)$ is the Von Neumann entropy~\cite{Note1}. For a
maximally-entangled resource state $\sigma_{AB}$ one has $C=2\log_{2}d$.

In a realistic scenario, noise must be explicitly included in the protocol.
For instance, noise can affect the transmission of quantum systems from the
sender (Bob) to the receiver (Alice), after the entangled resource state has
been perfectly distributed. This is the typical scenario in the definition of
entanglement-assisted protocols whose capacity is known~\cite{CEA1,CEA2}. More
realistically, noise may also affect the distribution itself of the resource
state from Alice to Bob. This scenario has been previously studied in
Refs.~\cite{noisy1, noisy2, RevMacchia} where it has been called
{\textquotedblleft two-sided\textquotedblright} noisy dense coding but not
capacity has been established.

This is the aim of this manuscript where the two-sided protocol is formulated
in a general feedback-assisted fashion. Here the round-trip transmission of
the quantum systems between Alice and Bob is interleaved by two adaptive
quantum operations (QOs) performed by Alice, which are optimized and updated
on the basis of the previous rounds.\ At the same time, Bob may also optimize
his classical encoding strategy, i.e., the probability distribution of his
Pauli encoders. Optimizing over these protocols we define the dense coding
capacity of a quantum channel between Alice and Bob. We then use simulation
techniques~\cite{PLOB,RevSim,Qmetro,revSENS} that allow us to simplify the
structure of the protocol and derive a single-letter upper bound for this
capacity. This quantity is explicitly computed for a Pauli channel in
arbitrary $d$ dimension, with remarkably simple formulas for qubit channels,
such as the depolarizing and the dephasing channel. \begin{figure}[ptbh]
\vspace{-0.5cm}
\par
\begin{center}
\includegraphics[width=0.43 \textwidth]{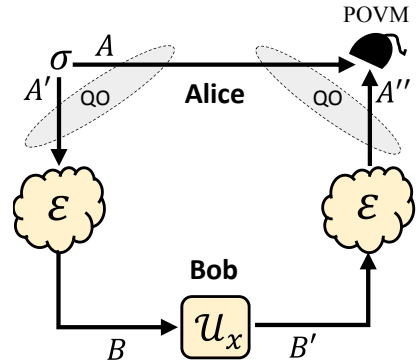} \vspace{-0.3cm}
\end{center}
\caption{Two-sided noisy dense coding over a quantum channel $\mathcal{E}$.
Alice prepares locally the bipartite state $\sigma_{AA^{\prime}}$ and sends
system $A^{\prime}$ to Bob who performs the Pauli unitary encoding
$\mathcal{U}_{x}$ and sends the system back through the quantum channel. At
the ouput Alice performs a joint positive-valued operator measure (POVM) in
order to retrieve $x$. In an adaptive version of the protocol, Alice performs
quantum operations (QOs) on her input and output systems which are generally
updated and optimized round-by-round. These QOs may also be conditioned by an
extra assisting variable which is communicated back by Bob.}%
\label{Fig:DenseProt}%
\end{figure}

\textit{Dense coding protocol}.--~Let us recall the expressions of Pauli
operators in a $d$-dimensional Hilbert space. On a computational basis
$\{|j\rangle\}$, we may define the two shift operators
\begin{equation}
X|j\rangle=|j\oplus1\rangle,~~Z|j\rangle=\omega^{j}|j\rangle, \label{shiftOP}%
\end{equation}
where $\oplus$ is modulo $d$ addition and $\omega:=\exp(2i\pi/d)$. We may then
consider the $d^{2}$ Pauli operators $X^{l}Z^{m}$ that, for simplicity, we
denote by $U_{x}$ with collapsed index $x=l,m$. For $d=2$, these operators
provide the standard qubit Pauli operators $X$, $Y$, $Z$ plus the identity
$I$. In the following we use the compact notation $\mathcal{U}_{x}%
(\rho):=U_{x}\rho U_{x}^{\dagger}$.

Now consider the scheme depicted in Fig.~\ref{Fig:DenseProt} where the
communication line between Alice and Bob is affected by a completely positive
trace-preserving (CPTP) map $\mathcal{E}$. Alice's resource state
$\sigma_{AA^{\prime}}$ is defined on a $d\times d$-dimensional Hilbert space.
Part $A^{\prime}$ is sent to Bob who encodes classical variable $X:=\{x,\pi
_{x}\}$ by means of $d^{2}$ Pauli operators $\mathcal{U}_{x}$ which are chosen
with probability $\pi_{x}$. In this way, Bob generates the state
\begin{equation}
\sigma_{AB^{\prime}}(x):=[\mathcal{I}_{A}\otimes(\mathcal{U}_{x}%
\circ\mathcal{E})_{A^{\prime}}](\sigma_{AA^{\prime}}),
\end{equation}
where $\mathcal{I}(\rho):=I\mathbb{\rho}I^{\dagger}$ is the identity map. Once
system $B^{\prime}$ is sent back through the channel, Alice receives the
output system $A^{\prime\prime}$ in the state $\rho_{AA^{\prime\prime}%
}(x):=(\mathcal{I}_{A}\otimes\mathcal{E}_{x})(\sigma_{AA^{\prime}})$ where we
have defined the encoding channel
\begin{equation}
\mathcal{E}_{x}:=\mathcal{E}\circ\mathcal{U}_{x}\circ\mathcal{E}.
\label{encodeX}%
\end{equation}

In order to retrieve the value of $x$, Alice performs a joint quantum
measurement on $A$ and $A^{\prime\prime}$. Asymptotically (i.e., for many
repetitions of the protocol), the accessible information of Alice's output
ensemble $\{\pi_{x},\rho_{AA^{\prime\prime}}(x)\}$ is given by the Holevo
bound~\cite{HOLE}
\begin{align}
\chi\left(  \{\pi_{x},\rho_{AA^{\prime\prime}}(x)\}\right)   &  =\\
&  S\left[
{\textstyle\sum_{x}}
\pi_{x}\rho_{AA^{\prime\prime}}(x)\right]  -%
{\textstyle\sum_{x}}
\pi_{x}S[\rho_{AA^{\prime\prime}}(x)].\nonumber
\end{align}
The one-shot dense coding capacity (1-DCC) of the channel $C_{D}%
^{(1)}(\mathcal{E})$ is obtained by optimizing over Bob's encoding variable
and Alice's input source, i.e., we may write
\begin{equation}
C_{D}^{(1)}(\mathcal{E})=\max_{\sigma,\pi_{x}}\chi\left(  \{\pi_{x}%
,\rho_{AA^{\prime\prime}}(x)\}\right)  \label{one-shotDCC}%
\end{equation}

\begin{figure}[ptbh]
\vspace{-1.5cm}
\par
\begin{center}
\includegraphics[width=0.48\textwidth]{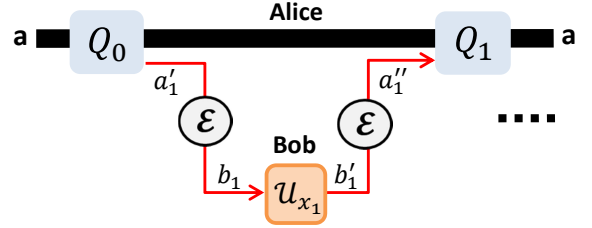} \vspace{-1.9cm}
\end{center}
\caption{Adaptive dense coding protocol over the quantum channel $\mathcal{E}%
$. Each encoding and round-trip transmission occurs between two quantum
operations (QOs). This is the first use of the protocol. See text for more
details.}%
\label{Fig:DenseAda}%
\end{figure}

\textit{Adaptive dense coding}{.--~Consider} now dense coding over a quantum
channel $\mathcal{E}$ where Alice performs QOs in an adaptive fashion. Alice
has a quantum register $\mathbf{a}$ as in Fig.~\ref{Fig:DenseAda}. At the
beginning Alice performs a QO $Q_{0}$ in order to prepare her initial state
$\rho_{\mathbf{a}}^{0}$. She then selects one system $a_{1}^{\prime}%
\in\mathbf{a}$ and sends it through $\mathcal{E}$. Once Bob receives the
output $b_{1}$, he encodes the message $x_{1}$ by means of a Pauli operator
$\mathcal{U}_{x_{1}}$ with probability $\pi_{x_{1}}$. This procedure gives
rise to the state $\rho_{\mathbf{a}b_{1}^{\prime}}^{0}(x_{1})$. Bob sends the
system $b_{1}^{\prime}$ backward to Alice through $\mathcal{E}$. At the
output, Alice incorporates the received system $a_{1}^{\prime\prime}$ in her
local register which is updated as $\mathbf{a}a_{1}^{\prime\prime}%
\rightarrow\mathbf{a}$. Next, Alice performs an optimized QO $Q_{1}$ on the
register with output state $\rho_{\mathbf{a}}^{1}(x_{1})$. In the second
transmission Alice picks another system $a_{2}^{\prime}\in\mathbf{a}$ and she
transmits it to Bob who receives $b_{2}$. Bob applies the second Pauli
operator $\mathcal{U}_{x_{2}}$ with probability $\pi_{x_{2}|x_{1}}$ and sends
the system back to Alice who performs another optimized QO $Q_{2}$ obtaining
the state $\rho_{\mathbf{a}}^{2}(x_{1}x_{2})$. After $n$ uses, Alice's the
output state will be $\rho_{\mathbf{a}}^{n}(\mathbf{x}_{n})$ where
$\mathbf{x}_{i}=x_{1}x_{2}\ldots x_{i}$ is the encoded message with
probability $\pi_{\mathbf{x}_{i}}=\pi_{x_{i}|x_{i-1}\cdots x_{1}}\cdots
\pi_{x_{2}|x_{1}}\pi_{x_{1}}$.

On average, Alice receives the ensemble $\{\pi_{\mathbf{x}_{n}},\rho
_{\mathbf{a}}^{n}(\mathbf{x}_{n})\}$ where the output state $\rho_{\mathbf{a}%
}^{n}(\mathbf{x}_{n})$ depends on the encoded classical information and the
sequence of QOs $\mathcal{Q}:=\{Q_{1},Q_{2},\ldots Q_{n}\}$. Alice's deferred
measurement~\cite{NiCh} will be done on the final state. For large $n$, and
optimizing the Holevo information of the ensemble $\mathcal{P}$ over all the
possible sequences $\mathcal{Q}$, we define the dense coding capacity (DCC) of
the quantum channel $\mathcal{E}$ as
\begin{equation}
C_{D}(\mathcal{E}):=\sup_{\mathcal{Q}}\max_{\pi_{\mathbf{x}_{n}}}\lim
_{n}n^{-1}\chi(\{\pi_{\mathbf{x}_{n}},\rho_{\mathbf{a}}^{n}(\mathbf{x}%
_{n})\})~. \label{DCC}%
\end{equation}
Note that this definition is more general than a regularized version
$C_{D}^{\infty}(\mathcal{E})$ of Eq.~(\ref{one-shotDCC}), where Alice prepares
a large multipartite input state, sends part of this state through $n$ uses of
the round-trip and then performal global measurement of the total output. In
fact, Eq.~(\ref{DCC}) assumes that Alice's input can also be updated
round-by-round on the basis of feedback from Bob~\cite{NotaC}.

\textit{Single-letter upper bound}.--~We now exploit a number of ingredients
from recent literature to derive a computable upper bound for the DCC. Recall
that, for any finite-dimensional quantum channel $\mathcal{E}$, we may write
the simulation $\mathcal{E}(\rho)=\mathcal{T}(\rho\otimes\sigma)$, where
$\mathcal{T}$\ is a trace-preserving LOCC and $\sigma$ a resource
state~\cite{PLOB}. Furthermore, suppose that the channel is covariant with
respect to Pauli operators so that, for any Pauli $\mathcal{U}$, we may write
$\mathcal{E}\circ\mathcal{U}=\mathcal{U}^{\prime}\circ\mathcal{E}$ for some
generally-different Pauli $\mathcal{U}^{\prime}$. In this case the channel is
Pauli-covariant and we may write~\cite{RevSim,PLOB} $\mathcal{E}%
(\rho)=\mathcal{T}_{\text{tele}}(\rho\otimes\sigma_{\mathcal{E}})$, where
$\mathcal{T}_{\text{tele}}$ is a teleportation LOCC and $\sigma_{\mathcal{E}}$
is the channel's Choi matrix, i.e., $\sigma_{\mathcal{E}}:=\mathcal{I}%
_{A}\otimes\mathcal{E}(\Phi_{AB})$ with $\Phi_{AB}$ being a
maximally-entangled state. Note that the Pauli unitaries $\mathcal{U}_{x}%
$\ are \textit{jointly} Pauli-covariant, i.e., we may certainly write
$\mathcal{U}_{x}\circ\mathcal{U}=\mathcal{U}^{\prime}\circ\mathcal{U}_{x}$
where $\mathcal{U}^{\prime}$ is the same for any $x$ (since a Pauli operator
either commutes or anti-commutes with another Pauli operator). Therefore, if
$\mathcal{E}$ is Pauli-covariant, we also have that the encoding channel
$\mathcal{E}_{x}$ is jointly Pauli-covariant. We may therefore write the
channel simulation $\mathcal{E}_{x}(\rho)=\mathcal{T}_{\text{tele}}%
(\rho\otimes\sigma_{\mathcal{E}_{x}})$ in terms of its Choi
matrix.
\begin{equation}
\sigma_{\mathcal{E}_{x}}:=\mathcal{I}_{A}\otimes\mathcal{E}_{x}(\Phi_{AB}).
\label{ChoiEx}%
\end{equation}
\begin{figure}[ptbh]
\vspace{-5.4cm}
\par
\begin{center}
\includegraphics[width=0.38\textwidth]{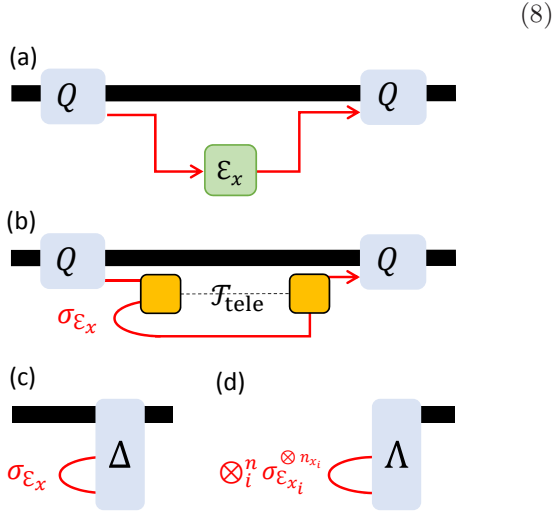} \vspace{+1.2cm}
\end{center}
\caption{Stretching of adaptive dense coding. (a)~Replace the round-trip
process of Fig.~\ref{Fig:DenseAda} with a total $x$-depending channel
$\mathcal{E}_{x}$. (b)~Simulate $\mathcal{E}_{x}$ by teleporting
$\mathcal{T}_{\text{tele}}$ over the Choi matrix $\sigma_{\mathcal{E}_{x}}$.
(c)~Collapse the QOs and the LOCC $\mathcal{T}_{\text{tele}}$ into a single QO
$\Delta$. (d) Repeat for all the $n$ uses, so as to collect an ensemble of
Choi matrices subject to a global QO $\Lambda$. }%
\label{Fig:stretch}%
\end{figure}

The next step is the stretching of the protocol as explain in
Fig.~\ref{Fig:stretch}. Thanks to this procedure the output state can be
decomposed in a tensor product of Choi matrices up to a global QO $\Lambda$,
i.e., we may write
\begin{equation}
\rho_{\mathbf{a}}^{n}(\mathbf{x}_{n})=\Lambda(\sigma_{\mathcal{E}_{x_{1}}%
}^{\otimes n_{x_{1}}}\otimes\sigma_{\mathcal{E}_{x_{2}}}^{\otimes n_{x_{2}}%
}\otimes\ldots\otimes\sigma_{\mathcal{E}_{x_{n}}}^{\otimes n_{x_{n}}})~,
\label{output}%
\end{equation}
where the $n_{x_{i}}$ is the number of $x_{i}$ occurrences in the message
$\mathbf{x}_{n}$. This given by $n_{x_{i}}=n\pi_{x_{i}}$ where $\pi_{x_{i}%
}=\sum_{j\neq i}\pi_{\mathbf{x}_{n}}$ is the marginal probability. Thanks to
Eq.~(\ref{output}) we can simplify the Holevo quantity in Eq.~(\ref{DCC}). In
fact, by using ($\star$)\ the contractivity under CPTP maps of the Holevo
quantity, and ($\bullet$) the subadditivity of the Von Neumann entropy $S$
under tensor products, we may write
\begin{align}
&  \chi(\{\pi_{\mathbf{x}_{n}},\rho_{\mathbf{a}}^{n}(\mathbf{x}_{n}%
)\})\overset{(\star)}{\leq}\chi\left(  \left\{  \pi_{\mathbf{x}_{n}},%
{\textstyle\bigotimes_{i=1}^{n}}
\sigma_{\mathcal{E}_{x_{i}}}^{\otimes n_{x_{i}}}\right\}  \right) \nonumber\\
&  =S\left(
{\textstyle\sum_{\mathbf{x}_{n}}}
\pi_{\mathbf{x}_{n}}%
{\textstyle\bigotimes_{i}}
\sigma_{\mathcal{E}_{x_{i}}}^{\otimes n_{x_{i}}}\right)  -%
{\textstyle\sum_{\mathbf{x}_{n}}}
\pi_{\mathbf{x}_{n}}S\left(
{\textstyle\bigotimes_{i}}
\sigma_{\mathcal{E}_{x_{i}}}^{\otimes n_{x_{i}}}\right) \nonumber\\
&  \overset{(\bullet)}{\leq}n_{x_{1}}S(%
{\textstyle\sum_{\mathbf{x}_{n}}}
\pi_{\mathbf{x}_{n}}\sigma_{\mathcal{E}_{x_{1}}})+\ldots+n_{x_{n}}S(%
{\textstyle\sum_{\mathbf{x}_{n}}}
\pi_{\mathbf{x}_{n}}\sigma_{\mathcal{E}_{x_{n}}})\nonumber\\
&  -n_{x_{1}}%
{\textstyle\sum_{\mathbf{x}_{n}}}
\pi_{\mathbf{x}_{n}}S\left(  \sigma_{\mathcal{E}_{x_{1}}}\right)
-\ldots-n_{x_{n}}%
{\textstyle\sum_{\mathbf{x}_{n}}}
\pi_{\mathbf{x}_{n}}S\left(  \sigma_{\mathcal{E}_{x_{n}}}\right) \nonumber\\
&  \leq nS\left(
{\textstyle\sum_{x}}
\pi_{x}\sigma_{\mathcal{E}_{x}}\right)  -n%
{\textstyle\sum_{x}}
\pi_{x}S\left(  \sigma_{\mathcal{E}_{x}}\right) \nonumber\\
&  =n\chi(\{\pi_{x},\sigma_{\mathcal{E}_{x}}\}), \label{stret2}%
\end{align}
where $\pi_{x}$ is the marginal probability of a generic letter $x$ and the
Choi matrix $\sigma_{\mathcal{E}_{x}}$ is defined in Eq.~(\ref{ChoiEx}). Note
that, in the last inequality of Eq.~(\ref{stret2}), we also use the fact that
a random code~\cite{NiCh,Shan}, i.e., a code where the codewords are randomly
chosen with an iid distribution equal to the marginal probability $\pi_{x}$,
is known to achieve the Holevo bound for discrete memoryless quantum
channels~\cite{HSW1,HSW2}.

By using Eq.~(\ref{stret2}) in the definition of Eq.~(\ref{DCC}), we may then
get rid of the supremum over $\mathcal{Q}$ and the asymptotic limit in $n$. We
may therefore write a single-letter upper bound for the DCC of a
Pauli-covariant channel $\mathcal{E}$ as
\begin{equation}
C_{D}(\mathcal{E})\leq\max_{\pi_{x}}\chi(\{\pi_{x},\sigma_{\mathcal{E}_{x}%
}\})~, \label{UB}%
\end{equation}
where $\pi_{x}$ is the marginal probability distribution of Bob's encoding
variable, and $\sigma_{\mathcal{E}_{x}}$ is the Choi matrix of the encoding
channel $\mathcal{E}_{x}$ in Eq.~(\ref{encodeX}). Note that the upper bound in
Eq.~(\ref{UB}) may be reached asymptotically by a non-adaptive protocol where
Alice prepares maximally-entangled states $\Phi_{AA^{\prime}}$ and sends
$A^{\prime}$ through the channel, while Bob applies independent Pauli
operators $\mathcal{U}_{x}$ with optimized probability $\pi_{x}$. Therefore,
for a Pauli-covariant channel we conclude that
\begin{equation}
C_{D}(\mathcal{E})=C_{D}^{(1)}(\mathcal{E})=\max_{\pi_{x}}\chi(\{\pi
_{x},\sigma_{\mathcal{E}_{x}}\}). \label{capDC}%
\end{equation}
Remarkably, no adaptiveness or regularization is needed to achieve the best
possible dense coding performance with a Pauli-covariant channel.

\textit{Dense coding capacity of Pauli channels}.--~The main result in
Eq.~(\ref{capDC}) can be applied to any Pauli channel at any finite dimension
$d.$ For any $d\geq2$, a Pauli channel takes the form
\begin{equation}
\mathcal{E}^{d}(\rho)=\sum_{k,r=0}^{d-1}p_{kr}\left(  X^{k}Z^{r}\right)
\rho\left(  X^{k}Z^{r}\right)  ^{\dagger},
\end{equation}
where $p_{kr}$ is a probability distribution, and $X$ and $Z$ are the
$d$-dimensional shift operators in Eq.~(\ref{shiftOP}). For this channel, we
may easily write an explicit formula for its DCC capacity. In evaluating the
Holevo bound, we notice that von Neumann entropy $S\left(
{\textstyle\sum_{x}}
\pi_{x}\sigma_{\mathcal{E}_{x}^{d}}\right)  $ is maximized by the uniform
probability $\pi_{x}=1/d^{2}$ and we can write $S\left(
{\textstyle\sum_{x}}
\pi_{x}\sigma_{\mathcal{E}_{x}^{d}}\right)  =\log_{2}d^{2}$. Then, using the
invariance of the entropy under unitary transformations, one has $%
{\textstyle\sum_{x}}
\pi_{x}S(\sigma_{\mathcal{E}_{x}^{d}})=S\left[  \mathcal{I}_{A}\otimes
\mathcal{E}^{d}(\sigma_{\mathcal{E}^{d}})\right]  $. Therefore, for the Holevo
quantity in Eq.~(\ref{capDC}) we may write
\begin{equation}
C_{D}(\mathcal{E}^{d})=\log d^{2}-S\left[  \mathcal{I}_{A}\otimes
\mathcal{E}^{d}(\sigma_{\mathcal{E}^{d}})\right]  .
\end{equation}
As expected this is strictly less than the entanglement-assisted classical
capacity of the channel, given by~\cite{CEA1,CEA2}
\begin{equation}
C_{E}(\mathcal{E})=\log d^{2}-S(\sigma_{\mathcal{E}^{d}})~.
\end{equation}

Consider a qubit depolazing channel, which is a Pauli channel of the form
\begin{equation}
\mathcal{E}_{\text{depol}}^{2}(\rho)=\left(  1-\frac{3}{4}p\right)  \rho
+\frac{p}{4}(X\rho X+Y\rho Y+Z\rho Z),
\end{equation}
for some probability $p$. Then, it is straighforward to see that
$C_{D}(\mathcal{E}_{\text{depol}}^{2})=2-h_{2}\left(  \alpha\right)
-\alpha\log3$, where $h_{2}(x)=-x\log x-(1-x)\log(1-x)$ is the binary entropy
function and $\alpha:=3/4p(2-p)$. Then, consider a qubit dephasing channel,
which takes the form
\begin{equation}
\mathcal{E}_{\text{deph}}^{2}(\rho)=(1-p)\rho+pZ\rho Z~.
\end{equation}
Its DCC is equal to the following expression
\begin{equation}
C_{D}(\mathcal{E}_{\text{deph}}^{2})=2[1-h_{2}(p)],
\end{equation}
for $p\leq1/2$ and zero otherwise.

\textit{Conclusion}.--~In this work we have considered the most general
adaptive protocol for the dense coding of classical information in a realistic
scenario where noise affects both the communication lines between Alice and
Bob. Assuming that this noise is modelled by the same quantum channel, we
define its dense coding capacity as the maximum amount of classical
information (per round-trip use) that Bob can transmit to Alice. We assume
that Bob is implementing Pauli encoders with an optimized probability
distribution and Alice is using quantum registers that are adaptively updated
and optimized in the process. For Pauli-covariant channel, we find that this
capacity reduces to a single-letter version based on a protocol which is
non-adaptive and one-shot (i.e., using iid input states). In particular, we
can establish exact formulas for the dense coding capacity of Pauli channels.

Note that our approach departs from the definition of entanglement-assisted
classical capacity of a quantum channel~\cite{CEA1,CEA2}, where it is
implicitly required that the parties either have a noiseless side quantum
channel for distributing entangled sources or they have previously met and
stored quantum entanglement in ideal long-life quantum memories. Our treatment
and definition of dense coding capacity removes these assumptions assuming
that the entanglement source is itself distributed through the noisy channel
and, therefore, it is realistically degraded by the environment. Because of
this feature, our capacity can also be seen as an upper bound for the key
rates of two-way quantum key distribution protocols that are related to the
dense coding idea~\cite{pp1,pp2,pp3,pp4,pp5}.

\textit{Acknowledgments}.--~This work was supported by the EPSRC via the `UK
Quantum Communications Hub' (EP/M013472/1) and the EC via \textquotedblleft
Continuous Variable Quantum Communications\textquotedblright\ (CiViQ, 820466).

\end{document}